\documentstyle{cupconf}

\ifoldfss
\else
  \ifnfssone
    \newmathalphabet{\mathit}
      \addtoversion{normal}{\mathit}{cmr}{m}{it}
      \addtoversion{bold}{\mathit}{cmr}{bx}{it}
    \newmathalphabet{\mathcal}
      \addtoversion{normal}{\mathcal}{cmsy}{m}{n}
    \else
    \ifnfsstwo
    \fi
  \fi
\fi

\def\hexnumber#1{\ifcase#1 0\or1\or2\or3\or4\or5\or6\or7\or8\or9\or
 A\or B\or C\or D\or E\or F\fi }

\makeatletter
\ifx\CUP@mtlplain@loaded\undefined
\else
\fi
\makeatother
 \makeatletter
 \ifx\CUP@mtlplain@loaded\undefined
   \font\tenbmi=cmmib10 at 10pt
   \font\sevenbmi=cmmib10 at 7pt
   \font\fivebmi=cmmib10 at 5pt

   \newfam\bmifam
   \textfont\bmifam=\tenbmi
   \scriptfont\bmifam=\sevenbmi
   \scriptscriptfont\bmifam=\fivebmi
   
 \fi
 \makeatother
\ifnfsstwo

\fi
\ifnfssone

\fi
\ifoldfss

\fi

\mathchardef\varLambda="0103

\makeatletter
\ifx\CUP@mtlplain@loaded\undefined
\else
\fi
\makeatother
\makeatletter
\ifx\CUP@mtlplain@loaded\undefined
  \font\tenbms=cmbsy10
  \font\sevenbms=cmbsy10 at 7pt
  \font\fivebms=cmbsy10 at 5pt
  \newfam\bmsfam
  \textfont\bmsfam=\tenbms
  \scriptfont\bmsfam=\sevenbms
  \scriptscriptfont\bmsfam=\fivebms

  \edef\bsy@{\hexnumber\bmsfam}
  \mathchardef\bnabla="0\bsy@72
\fi
\makeatother
\title[Initial Stellar Mass Function]
{Observation and Theory of The Initial Stellar Mass Function
\footnote{
to be published in "Unsolved Problems in Stellar Evolution," 
ed. M. Livio, Cambridge University Press, 1998}
}
\author[B. G. Elmegreen]{%
B\ls R\ls U\ls C\ls E\ls \ns
G.\ns
E\ls L\ls M\ls E\ls G\ls R\ls E\ls E\ls N}
\affiliation{IBM Research Division, T.J. Watson Research Center,
P.O. Box 218, Yorktown Hts NY 10598 USA bge@watson.ibm.com}

\begin{document}
\ifnfssone
\else
  \ifnfsstwo
  \else
    \ifoldfss
      \let\mathcal\cal
      \let\mathrm\rm
      \let\mathsf\sf
    \fi
  \fi
\fi

\maketitle

\begin{abstract} Observations of normal galactic star-forming regions
suggest there is widespread near-uniformity in the initial stellar mass
function (IMF) in spite of diverse physical conditions. Fluctuations
may come largely from statistical effects and observational selection.
There are also tantalizing, but uncertain reports that
the IMF shifts systematically in peculiar regions,  giving a low
mass bias in quiescent gas, and a high mass bias in active
starbursts.  Theoretical proposals for the origin of the IMF are
reviewed.  The theories generally focus on a
combination of four physical effects: wind-limited accretion of stellar
mass, coalescence of protostellar gas clumps, mass limitations at the
thermal Jeans mass, and power-law cloud structure.  Hybrid theories
combining the best of each may be preferred.
\end{abstract}    

to be published in "Unsolved Problems in Stellar Evolution,"
ed. M. Livio, Cambridge University Press, 1999, in press.

\firstsection

\section{Observations}

\subsection{Hints at a Universal IMF in Normal Star-Forming Regions}
\label{sect:intro}

The initial stellar mass function (IMF) has been observed directly in
dozens of clusters in our Galaxy and the Large Magellanic Clouds,
and inferred indirectly for our Galaxy and other galaxies from
metallicities, supernova rates, ionization levels, luminosities,
dynamical masses, colors, and other tracers.

There is some convergence in these observations to an IMF that is
approximately a power law above 1 M$_\odot$, having a slope between
$x\sim-1$ and $\sim-1.6$ on a $\log N -\log M$ histogram (see reviews in
Massey 1998 and Scalo 1998), with a flattening below this down to
$\sim0.1$ M$_\odot$ or lower. The power-law part is in rough agreement
with the Salpeter (1955) function, which has a slope of $x=-1.35$ on such
a plot, and also with the Scalo (1986) determination, which revised the
often-used Miller \& Scalo (1979) function at intermediate and high mass
(the Miller-Scalo function is too steep). 

Scalo (1998) recently emphasized the variations in the derived power law
slopes from region to region, which amount to about $\pm0.5$. There are
no obvious correlations between these variations and physical properties
of the star-forming regions, including the cluster density,
galactocentric radius, and metallicity (Scalo 1998). There are also
``gaps'' in the power law parts of some cluster IMFs, according to Scalo
(e.g., NGC 103; Phelps \& Janes 1993), and a ``ledge'' in the Orion IMF
by Hillenbrand (1997). These features led Scalo to suggest that physical
effects might be involved, making IMF variations larger than Poisson
statistics. 

An emphasis on the similarities, rather than the differences, in the IMF
has been made by Massey and collaborators (see review in Massey
1998). They find an IMF power law slope that is fairly constant, around
the Salpeter value, for clusters with a factor of $\sim200$ range in
densities (Massey \& Hunter 1998; Luhman \& Rieke 1998) 
and a factor of $\sim10$ range in
metallicities (Massey, Johnson \& DeGioia-Eastwood 1995a; see also the
globular cluster study by Bellazzini et al. 1995). This slope was also
found by Kennicutt, Tamblyn \& Congdon (1994) in galaxy-wide surveys,
and by Bresolin \& Kennicutt (1997) for individual HII regions in
galaxies, using the equivalent widths of hydrogen emission lines, and it
was found by Greggio et al. (1993), Marconi et al. (1995), Holtzman, et al.
(1997), and Grillmair, et al. (1998) from color magnitude diagrams of
stars in the LMC and local dwarf galaxies. 

Chemical evolution also tends to suggest an invariant IMF, at least at
intermediate and high mass, in the stellar halo (Nissen et al. 1994) and
disk (Tsujimoto et al. 1997; Wyse 1998), in QSO damped Ly$\alpha$
(Lu et al. 1996) and Ly$\alpha$ forest (Wyse 1998) lines, the
intracluster medium (Renzini et al. 1993; Wyse 1997, 1998; however, see
Loewenstein \& Mushotzky 1996), and elliptical galaxies (see review in
Wyse 1998). The observable in this case is the ratio of the iron to the
oxygen abundance, which is always about the same. Since this ratio
scales with the ratio of intermediate mass to high mass stars, the IMF
must be about the same too. A different result was obtained by
Heikkil\"a, Johansson \& Olofsson (1998), who conjectured that a low
C$^{18}$O/C$^{17}$O ratio in the LMC results from a relatively low
proportion of high-mass stars, as observed, for example, by Massey et
al. (1995b); clearly the chemical and fractionation processes involved
with this important isotope ratio have to be investigated further.

Gibson \& Matteucci (1997) also obtained a different result from
evolution and metallicity models of elliptical galaxies, including
infall and wind ejection of gas.  They suggested the IMF had a higher
proportion of massive stars than the Solar neighborhood, as it appears
to in some starburst models (discussed below).  Angeletti \& Giannone
(1997) modeled elliptical galaxies and fit the mass-to-light and [Fe/H]
ratios with an IMF that has a mass-dependent high mass slope, different
from the Salpeter value.

The IMF in local OB associations might differ from the Salpeter value
too.  Brown (1998) reviewed observations, including those with 
distances from the Hipparcos
satellite, which suggest a slightly steeper slope, $\sim-1.7$ to $-1.9,$
for local associations, 
but the stellar types in these studies were obtained from photometry, 
and Massey (1998)
discussed how such data can artificially steepen the IMF
without spectroscopic observations too. 

At the low mass end, there is no {\it systematic} indication yet of a
turnover at masses lower than 0.1 M$_\odot$, but there are some regions
where such a turnover has been reported, e.g., in nearby field stars
(Reid \& Gazis 1997a), Orion (Hillenbrand 1997), the Pleiades (Reid
1998), NGC 6231 (Sung, Bessell, \& Lee 1998), 30 Dor (Nota 1998), and
the globular cluster NGC 6397 (King et al. 1998). There are also
regions where the IMF appears to rise below 0.1 M$_\odot$, again for
local stars (M\'era et al.  1996; Gould et al. 1997). Scalo (1998)
suggests that the local field results at low mass contain uncertainties
in the star formation rate, pre-main-sequence contribution, and
mass-luminosity relation. The older cluster results are uncertain also
because of evaporation of low mass members, although Reid (1998)
suggests evaporation is not important for the Pleiades. In any case,
Scalo (1998) and Reid (1998) both provide evidence that the low mass
IMF, below 1 M$_\odot$, varies slightly in shape from region to region,
although no sensible correlations with properties of the environment
have yet been noticed. Variations in the low mass IMF are also seen for
globular clusters, without any obvious dependence on metallicity
(Santiago et al. 1996; Piotto et al. 1997; see review in Cool 1998),
but there may be some dependence on the cluster evaporation rate and
susceptibility to orbit shocking (Cool 1998), since low mass stars
leave the cluster most easily. The IMF for the young open cluster
mentioned above, NGC 6231 (Sung et al. 1998), is somewhat peculiar,
because the turnover mass is around 2.5 M$_\odot$, much higher than in
OB associations and other clusters. However, NGC 6231 is a dense
cluster $\sim10$ My old, and a high turnover mass in the central
regions might be expected from rapid thermalization leading to mass
segregation (e.g., Scalo 1986).

The low mass IMF is also probed by observations of embedded clusters,
where most of the stars are pre-main sequence (Zinnecker et al. 1993;
Fletcher \& Stahler 1994a,b; Lada \& Lada 1995; Lada, Alves \& Lada
1996; Megeath 1996; Horner, Lada ,\& Lada 1997). 
Pre-main-sequence (PMS) clusters have not yet
evolved kinematically, so differential evaporation and segregation of
high and low mass members should not be as much of a problem as it is
for older clusters. PMS stars are also much brighter than they will be
on the main sequence, so even stars less massive than the hydrogen
burning limit can be observed in local clusters (see review in Lada,
Lada, \& Muench 1998). Lada et al. (1998) combined Herbig's (1998)
optical data on the cluster IC 348, which was used to determine the
history of star formation in the region, with their own K-band
luminosity function to determine the low mass IMF. They found that the
IMF in this cluster turns over below 0.25 M$_\odot$.

Recent searches for brown dwarfs (Zuckerman \& Becklin 1992; Reid \&
Gizis 1997b; Reid 1998) suggest these objects are so infrequent that
the IMF cannot continue to rise at low mass. The number of possible
brown dwarfs is consistent with an extension of the flattening seen
between 1 M$_\odot$ and 0.1 M$_\odot$ (Reid 1998).

\subsection{Starburst IMFs and the Extreme Field}

In spite of these approximate similarities in the IMF from region to
region, there have been two clear and consistent deviations that, if
real, would have important consequences for star formation. One is the
apparent shift toward relatively higher masses in regions of extreme
activity, such as starburst galaxies (e.g. Rieke et al. 1980), and the
other is the apparent shift toward relatively lower masses in regions
of extreme {\it in}activity, the so-called ``field'' population, away
from OB associations, as discussed by Massey et al. (1995b). Indeed,
there may be a monotonic transition from low-mass weighted IMFs to
high-mass weighted IMFs with increasing star formation rate, ISM
pressure, radiation field, etc., perhaps because of a change in cloud
properties, cloud destruction capabilities of young stars, and other
things. The ``normal'' IMF occupies what seems to be a broad range of
star formation rates between these two limits, but this broad range is
not so broad in terms of {\it possible} ISM conditions. It may only be
broad in the sense that it is common in main galaxy disks where some
type of feedback between star formation and gas stability is
maintained for long periods of time.

Reviews of IMFs in starburst galaxies are in Scalo (1990), Zinnecker
(1996), and Leitherer (1998). The first observation that suggested a
``top-heavy'' IMF was that the luminosity of the starburst regions are
so high, and the dynamical masses from the rotation curves so low, that
there cannot be a full complement of low mass stars, which dominate the
mass, to go with the high mass stars, which dominate the light (Rieke
et al. 1980, 1993; Kronberg, Biermann, \& Schwab 1985; Wright et al.
1988; see reviews in Telesco 1988, Joseph 1991, Rieke 1991, Thuan
1991).  Thus the IMF contains a higher proportion of high mass stars
than the Solar neighborhood.  Doane \& Matthews (1993) reached the same
conclusion for M82 based on many properties of galactic evolution
models, including the remaining gas mass, total mass, star formation
rate, and supernova rate.  Doyon, Joseph, \& Wright (1994) and Smith et
al. (1995) found top-heavy IMFs from spectroscopic line ratios for the
starburst galaxies NGC 3256 and UGC 8387, respectively.  Smith, Herter
\& Haynes (1998) found the same result from infrared excesses in 20
starburst galaxies.

The distribution of stars was often not resolved in the early starburst
studies. Satyapal et al. (1995, 1997) derived a smaller
extinction-corrected K band luminosity for the inner 30'' of the
starburst galaxy M82, and suggested the IMF could have the Salpeter
slope all the way from 0.1 to 100 M$_\odot$ with only 30\% of the
dynamical mass in the star formation event.  Ho \& Filippenko (1996a,b)
measured the velocity dispersions in two dense clusters in starburst
galaxies, clusters that may evolve into globulars like those in the
halo of our Galaxy (Meurer 1995). They found that the young globulars
have large masses like old globulars, and that the young clusters would
evolve into old clusters with a mass-to-light ratio like the old
globulars after a Hubble time.  Because of this, they suggested that
the IMF might be the same in the young starburst globulars as it is in
old globular clusters, and in particular, that it contains low mass
stars (e.g., De Marchi \& Paresce 1997).

Other considerations have led to revisions of the top-heavy models as
well.  Devereux (1989) observed 20 nearby starbursts like M82 with
infrared photometry, and found no evidence for an unusual IMF in any of
these; he pointed out that a revision of extinction corrections for M82
could change the Rieke et al. (1980) result too, as recently confirmed
by Satyapal et al. (1997).  Schaerer (1996) considered detailed
evolutionary models of young stellar populations in starburst galaxies
and also found that the IMF was normal.  Calzetti (1997) modeled
multiwavelength spectroscopy and broad-band infrared photometry of 19
starburst galaxies and found that the Salpeter IMF between 0.1 and 100
M$_\odot$ fits the observations.  Stasi\'nska \& Leitherer (1996)
modeled the emission line spectra of giant HII region and starburst
galaxies, having a factor-of-ten range in metallicities, and found a
Salpeter IMF up to 100 M$_\odot$; the lower mass
limit could not be obtained, however.

The proposed peculiarity of starburst IMFs is therefore not certain, a
conclusion reached by Scalo (1990) as well. Maybe only the extreme
examples of starbursts require a high proportion of massive stars,
while other starbursts, perhaps more ``ordinary,'' do not. The exact
form of the peculiarity in a starburst IMF is not known either. It
could be ``truncated'' in the sense that the lower mass limit of the
power law, where the power law slope begins to flatten, is higher than
usual, perhaps as high as several M$_\odot$ instead of several tenths
of a M$_\odot$. Such truncation may be observed already for 30 Dor
(Nota 1998). Or, it could have a power-law slope that is shallower than
$-1.35$, or possibly, a higher upper mass limit. These differences are
often hard to distinguish, even though all of them affect the
mass-to-light ratio. Fortunately, the exact form of the IMF sometimes
affects other things as well. For example, Charlot et al. (1993)
suggested that a truncated IMF will produce a very red population of
red giants, without the corresponding main sequence stars, after the
turnoff age reaches the stellar lifetime at the truncation mass. Wang
\& Silk (1993) suggested that the truncated model gives an oxygen
abundance that is too high when the star formation process is over.
Presumably these other constraints will help narrow down the possible
choices.

The extreme field studied by Massey et al. (1995b) is another example
where the Salpeter IMF seems to break down. In regions of the LMC and
our Galaxy that are more than 30 parsecs from catalogued OB
associations, Massey et al. found O-type stars that probably did not
drift there during their short lifetimes. Massey et al. showed with an
H-R diagram that there was an approximately steady rate of star
formation there for the last 10 My years, and they derived a
time-averaged IMF. The result was very steep, with $x\sim4$ instead of
$\sim1.35$ for a Salpeter function.

\subsection{Other IMF oddities}

There are several other peculiarities of the IMF that may or may not
have a specific physical origin. For example, it has been known for a
long time (Larson 1982; Myers \& Fuller 1993) that high mass clouds
produce high mass stars, and low mass clouds produce low mass stars.
Two decades ago, this observation and others were related to the
concept of a bimodel IMF (Eggen 1976, 1977; Elmegreen 1978; G\"usten \&
Mezger 1983; Larson 1977, 1986), and to the impression that spiral arms
preferentially formed massive stars (Mezger \& Smith 1977), presumably
because the molecular clouds in spiral arms are more massive, or
somehow different from interarm clouds. Today, we know that massive
clouds also produce low mass stars in great abundance (Herbig \&
Terndrup 1986), and there are essentially no local regions that produce
high mass stars exclusively. Thus the ``observation'' that the maximum
mass of a star correlates with the mass of the cloud could be only a
statistical sampling effect (Larson 1982; Elmegreen 1983; Walter \& Boyd
1991; Massey \&
Hunter 1998): high mass clouds produce more stars of all types, and so
they are more likely to sample at the tail of the IMF to produce a few
massive stars. In the same sense, high mass clouds should also produce
more dwarf stars than low mass clouds.

\begin{figure}
\vspace {2.6in}
\includegraphics{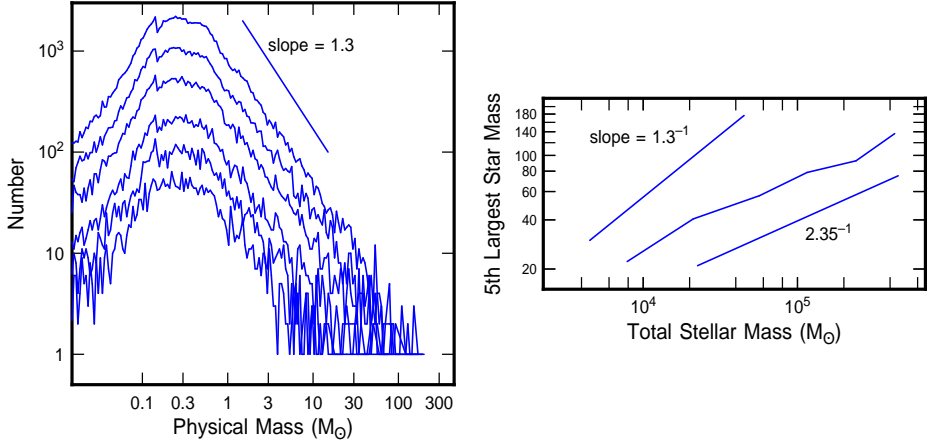}
\caption{Model IMFs with different numbers of stars
showing an increase in the
maximum stellar mass with increasing number of stars.
The figure on the right shows the mass of the fifth largest star
versus the total stellar mass, with an expected slope equal to $1/x=1/1.3$
and a measured slope equal to $\sim1/2.35$.  }
\label{fig:mvsm}
\end{figure}

An IMF model based on random sampling from a hierarchical cloud
(Elmegreen 1997a) shows this effect clearly in figure \ref{fig:mvsm}.
Each curve on the left is from a different cluster, with a number of
stars in the clusters equal to 2500,
5000, 10000, 25000, 50000, and 100000.
As the number of model
stars increases, the mass of the largest star increases too, as shown
on the right of the figure.  The expected slope of the power law
relation between cluster mass, $M_{cl}$, and maximum star mass,
$M_{max}$, is $1/x$, i.e., $M_{max}\sim M_{cl}^{1/x}$; actually the
calculation gives a slightly steeper slope because of storage
limitations to the number of hierarchical levels allowed in the
simulation (these models use 8 levels with an average of 3 subclumps
per clump at each level).  This property of increasing maximum star
mass with cloud mass should not depend on any specific IMF model or
on the star formation process.

To be more precise about the stochastic interpretation of this effect,
we can consider an IMF $n(M)dM= n_0 M^{-1-x}dM$, which gives the
Salpeter function when $x=1.35$. The largest star, with mass $M_{max}$,
typically satisfies $\int_{M_{max}} ^\infty n(M)dM=1$.  This gives
$n_0=xM_{max}^x$.  Because the total cluster mass is
$M_{tot}=\int_{M_{min}}^{M_{max}}Mn(M)dM$, the largest star is related
to the total cluster mass as \begin{equation}
M_{tot}=
{{x}\over{x-1}}M_{max}\left[\left({{M_{max}}\over{M_{min}}}\right)^{x-1}
-1\right].  \end{equation} 
For $M_{max}\sim10 M_{min}$ and $x=1.35$,
this gives $M_{tot}\sim4.8M_{max}$, and for
$\left(M_{max}/M_{min}\right)^{0.35}$ $>>1$, this gives $M_{tot}\sim
3.9M_{max}^{1.35}/M_{min}^{0.35}$.  Thus, a cluster requires several
hundred M$_\odot$ before a small O-type star is likely to appear, and a
cluster with $10^3$ M$_\odot$ is likely to contain $\sim10$ modest
O-type stars or a few very massive ones.  This is the mass range for
the largest open clusters in the Milky Way disk, indicating that when a
few O stars appear, a cluster is likely to terminate its formation in
an unbound state, presumably as a result of rapid cloud destruction,
and make an expanding OB association (Elmegreen 1983).

Larson (1991, 1992) and Khersonsky (1997) interpret this observation in
the opposite way, saying that there is a physical effect which
tends to produce higher mass stars in clouds of higher mass, and that
the IMF partly results from this physical effect.  Larson (1992)
attributes the physical effect to a larger accretion length in larger
clouds for material going into a star.  If there is such an effect, and
the observation is not entirely statistical as discussed above, then
the superposition of IMFs from clouds of different mass would be
different from the IMF in any one cloud. That is, clouds of low mass
would produce an IMF with a particular slope up to some maximum mass,
and clouds of higher mass would produce an IMF, perhaps with the
same slope, up to higher masses. The sum of these two IMFs would
be an IMF with greater slope than either separate cloud, because the
summed IMF contains low mass stars from both clouds, but only high mass
stars from the more massive cloud.  Larson (1991) proposed his model by
saying, in effect, that only one star forms in each cloud fragment at
each level in the hierarchy, and that the star mass is always
proportional to the square root of the fragment mass.  Then the sum
over all fragments is the final IMF, and this has a slope of $x=2$,
which is not too bad.  Stated this way, there is no problem with the
model because each component in the sum is only a single star, not a
whole cluster of stars with a separate IMF.  However, even in this
model, if we add together the stellar populations from many separate
clouds with different total masses, then the summed IMF will be steeper
than the IMF in each.  At this point, it is important to recall that
the IMF obtained from large-scale surveys of galaxies and their
metallicities is about the same as the IMF observed in individual
clusters.  Thus the summed IMF from many regions of star formation has
to be about the same as the IMF from each separate cluster.  This would
seem to suggest that the observed increase in maximum star mass with
cloud mass is not the result of a physical process that specifically
increases the upper stellar mass in larger clouds.

A second peculiarity about star formation is that the high mass stars
tend to form after the low mass stars (Herbig 1962a,b; Iben \& Talbot
1966).  This has been determined most recently for 30 Dor (Massey \&
Hunter 1998).  Again one can think of physical reasons for this, such
as a gradual warming of the cloud following the formation of low mass
stars, and an increase in the Jeans mass with temperature (Silk 1977;
Yoshii \& Saio 1985), but statistical effects can explain it too.

The decreasing nature of the IMF at high mass implies that massive
stars are likely to form only after a lot of low mass stars have
already formed.  For a constant star formation rate, the average time
between the formation of stars in a logarithmic mass range centered on
$M$ is proportional to the rate at which these stars form, which is
inversely proportional to the relative number of the stars, or
$\left(Mn[M]\right)^{-1}\propto M^{x}$.  For a constant star formation
rate, this is also the average time after star formation begins for the
{\it first} appearance of a star with this mass.  Thus the most common,
low-mass stars form first, followed by the intermediate and then the
high mass stars. In terms of the proportion of all stars formed, and
for a constant star formation rate in units of mass per year, the
maximum stellar mass increases with time as \begin{equation}
\int_{M_{min}}^{M_{max}(t)}Mn(M)dM=At\end{equation} for star formation
rate $A$.  Thus, a star of mass $M(t)$ forms after the proportional
time given by \begin{equation}
{{t}\over{t_{max}}}=
{{\int_{M_{min}}^{M(t)}Mn(M)dM}\over{\int_{M_{min}}^{M_{max}}Mn(M)dM}}.
\end{equation}
For the power law portion of the IMF only, with $M_{min}=1$ M$_\odot$
and $x=1.35$, this becomes, for masses in M$_\odot$,
\begin{equation}
{{t}\over{t_{max}}}={{1-M(t)^{-0.35}}\over{1-M_{max}^{-0.35}}}.
\end{equation}
For example, in a cluster with a maximum stellar mass of 30 $M_\odot$,
stars with 10 M$_\odot$ form in the last 20\% of the time. 

Note that this increase in time with stellar mass is only for the 
first appearance of a star with that mass. The {\it average} time of
appearance is independent of stellar mass in the stochastic interpretation.
This suggests a way to check the stochastic model of birth orders. 
If the average time of appearance of all stars with mass $M$ is
the average age of the cloud, and this is true for any $M$, 
and if the first time of appearance of a star with mass $M$ increases
with time, as described above, then the birth order is dominated
by stochastic effects. 

\begin{figure}
\vspace {2.6in}
\includegraphics{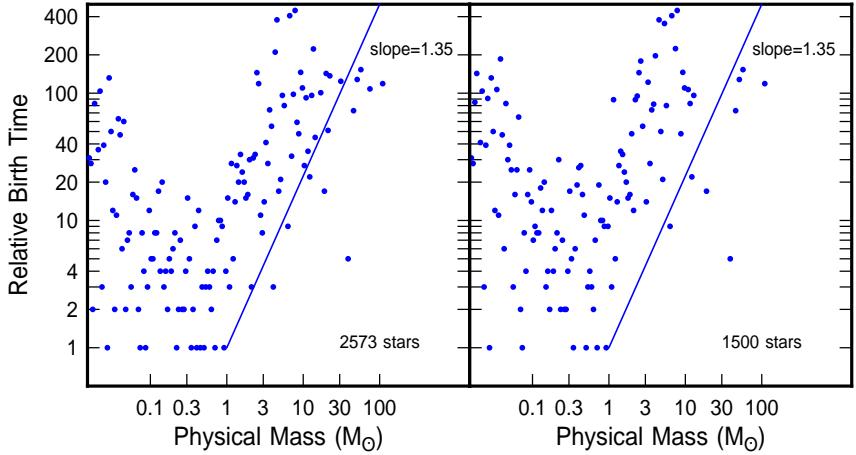}
\caption{The relative birth time for the first appearance of
a star with a certain mass is shown as a function of
mass for a composite of 11 clusters on the left, containing
2573 stars total, and 6 clusters on the
right, containing 1500 stars. The relative birth time is measured
in units of the time of first appearance of any star in the model.
The solid line has a slope of 1.35.
}
\label{fig:mt}
\end{figure}

Figure \ref{fig:mt} shows the relative birth order of stars in the
random sampling IMF model by Elmegreen (1998).  Each point marks the
earliest time of formation of a model star with that mass, among all of
the stars that formed in 11 clusters on the left and 6 clusters on the
right. Each cluster contains 200 to 500 stars and is independently
generated.  The plot shows a distribution of points that traces the
inverse of the IMF in log intervals. The fiducial line has a slope of
$1.35$ for comparison.  The increase at low mass occurs because the
model uses a Gaussian probability for failure to form a star at low
mass, which is an arbitrary assumption at this time (it causes the
model IMF to turn over at low mass).

There are no quantitative measures yet of how the earliest appearance
of a star with a particular mass increases with time relative to the
birth order of all stars, so the predictions of theory cannot yet be
checked.  In a real star-forming region, the star formation rate
is probably
not constant in time, so the ordinate in figure \ref{fig:mt} should not
be time, but star number in its proper birth order; i.e. one should
plot, for the first appearance of a star of mass $M$, its mass
versus the number of
stars that formed before it.

An interesting implication of the above two discussions is that {\it
massive stars must not be able to severely limit or halt star
formation in their primordial clouds.}  If they did, then, considering
their relatively late appearance, each cloud would produce stars with
the same local IMF up to some maximum stellar mass at which point the cloud
is finally destroyed. This maximum stellar mass at the time of cloud
destruction is presumably larger for more massive clouds, because it is
harder to destroy a massive cloud than a low mass cloud.  Thus the
summed IMF from many regions of star formation would be steeper than
the local IMF in each region, for the reasons discussed above, and we
could not explain why the integral IMF in a galaxy is the same as the
cluster IMF.  Perhaps massive stars only shred the low density
parts of these clouds, allowing (or perhaps stimulating),
most of the cores that have already formed
in these regions to continue their evolution
toward stars. In view of this, it seems that 
stars of all mass are probably equally
likely to form in any cloud (that contains at least this much mass),
regardless of what other stars have formed before them. Moreover, 
the maximum stellar mass is not physically limited, it is only
statistically limited in the sense that extremely massive stars are
rare.  An exception to this conclusion may arise in the ``field''
population studied by Massey et al. (1995b), in which the summed IMF of
presumed dispersed clusters appears to be steeper than the IMF in
each.  Maybe massive stars can more easily disrupt their clouds in low
pressure ``field'' environments. On the other hand, maybe the Massey et
al.  observation can be explained in other ways.

A third peculiarity of the IMF is that massive stars are often closer
to a cluster center than low mass stars.  This effect has been
discussed for a long time (e.g., Sagar et al. 1988), 
but it was never certain whether the
observations reflected a birth position, or just a relaxation effect
after the cluster formed. Thermal relaxation makes the high mass stars
sink to the center of the cluster, on average, in a thermalization time
scale. However, Hillenbrand \& Hartmann (1998) and Bonnell \& Davies
(1998) have recently shown that this peculiar mass distribution, which
is observed in the Trapezium cluster, could not result from
thermalization because the Orion cluster is too young (see also Jones
\& Walker 1988).  Similar mass segregation has been observed in 
NGC 2157 (Fischer et al. 1998).  
Thus it is an effect that should be explained by an
IMF theory, and it is likely to result from a physical process during
star formation rather than stochastic sampling.

There have been many attempts to explain this relative birth position
(see reviews in Larson 1991, and Zinnecker, McCaughrean \& Wilking
1993), including excess gas clump collisions in the center of a cloud
(Larson 1990; Stahler, Palla, \& Ho 1998), excess gas accretion onto
protostars in the cloud core (Larson 1978; Larson 1982; 
Zinnecker 1982; Bonnell et
al. 1997), enhanced gas drag for massive protostars (Larson 1990,
1991), late-time formation of high mass stars in a collapsing cloud
(Murray \& Lin 1996), and hierarchical cloud structure in which the
massive trunk of a hierarchical tree for the gas cloud is closer to the
center of the tree than the low-mass branches (Elmegreen 1998).  A
combination of these effects might be involved.

\subsection{Summary}
\label{sect:sum1}

There are evidently four distinct physical effects that have to 
be explained by a theory of the IMF:\\
(1) the power-law slope at intermediate to
high stellar mass, with a value close to that found by Salpeter (1955);\\
(2) the flattening at low mass for clusters and star-forming regions in
our Galaxy;\\
(3) the increase in the transition mass between power-law
and flat regions of the IMF in starburst galaxies,\\
(4)  the
preferential birth of high mass stars close to cluster centers.

There are apparently three additional observations
that can be explained by stochastic sampling effects:\\
(1) the tendency for the largest mass star in a
region to increase with the total number of stars;\\
(2)  the tendency for
the largest mass stars to form last, and\\
(3) the seemingly random
variations in the intermediate and high mass IMF slopes from region to
region, along with the appearance of gaps, ledges, and other
peculiar departures from a power law.

In addition to these observations, there are tentative indications that
other IMF features may be present, including:\\
(1) a turnover at low mass, and\\
(2) a steepening of the power-law part in field regions.\\
Future observations should be able to illuminate these uncertain features. 

\section{What Determines a Star's Mass? }
\subsection{A Steady Stream of Theories}
\label{sect:theory}

The theory of the IMF depends on both the theory of star formation and
the theory of cloud structure.  Because of the complexity of these
issues, the IMF may not be understood for a long time.  Here we review
some recent developments. Previous reviews were in Cayrel (1990) and
Clarke (1998).

One of the key problems in understanding the origin of the IMF is to
determine the processes that limit the pre-stellar gas accretion onto a
star and define the star's final mass.  There have been several ideas
on this.

The final mass could result from the star's own ability to limit the
accretion of new gas onto its surface, perhaps because of an intense
proto-stellar wind (Larson 1982; 
Shu, Adams, \& Lizano 1987), which is known to
exist at this phase of evolution (Lada 1985).  This possibility has led
to several theories of the IMF. One, proposed by Nakano, Hasegawa, \&
Norman (1995), considered three mass-limiting agents: protostellar
winds, ionization, and depletion of the gas clump in which the star
forms. They concluded that under normal conditions, the protostellar
wind would limit the accretion.   Then they calculated the fraction of
the clump that would get into the star, finding that the final star
mass scaled with the clump mass to the $7/6$ power.  To get an IMF
slope of $x=1.7$, which they got from the field-star IMF in Scalo
(1986), they required a very steep clump mass spectrum,
$n(M_{clump})\sim M_{clump}^{-3}$ in linear intervals, instead of the
usual observation of $M_{clump}^{-\alpha}$ with $\alpha\sim1.5-1.8$
(Blitz 1993; Kramer et al. 1998).  
They justified this steep clump spectrum by noting that
if the linewidths found in a CS survey of Orion clumps (Tatematsu et
al. 1993) were assumed to be virial velocities, and the masses
calculated accordingly, then the Orion CS clumps would have such a
spectrum.

Another IMF model with wind-limited stellar masses was proposed by
Adams \& Fatuzzo (1996).  They conjectured that the stellar luminosity
scales with the wind mass loss rate, which, at the time when the
accretion stops, is proportional to the direct accretion rate onto the
star, independent of what goes onto the disk.  This gave them a
relation between the stellar properties, i.e., luminosity and mass, and the
gas cloud properties, sound speed and angular rotation rate.  The
angular rotation rate entered because they had to determine the
fraction of the accreting gas that goes into the star and not the
disk.  Finally, they related the stellar mass to the total luminosity
of the star, from the sum of the accretion-driven luminosity, which
depends on mass, and the luminosity of a main sequence star at
intermediate mass, from the standard mass-luminosity relation.  The
result is a relation between stellar mass and the cloud properties,
i.e., sound speed and rotation rate.  Taking the mass distribution for
clumps with the observed slope $\alpha$ discussed above, and the
empirical relationship between clump mass and velocity dispersion from
Larson's (1981) laws, they got a distribution function for the clump
velocity dispersion, which then led to a distribution function for the
final star mass if the angular rate of all the clumps is the same.
Adams \& Fatuzzo (1996) also discussed more general equations giving
the stellar mass from cloud properties, considering random variations in
these cloud properties, and derived a log-normal final mass
distribution, as in Larson (1973), Elmegreen \& Mathieu (1983),
Zinnecker (1984), and Elmegreen (1985).

Silk (1995) also got a relation between cloud core mass and turbulent
linewidth, different from Adams \& Fatuzzo's, considering centrifugally
supported cores with luminosities equal to the accretion luminosities.
He considered an equality between the turbulent linewidth and the
thermal velocity in the core, which entered into the luminosity through
the temperature using the usual radiative transfer equation for a star
and the Rossland mean opacity. Then, after substituting the empirical
correlations between clump rotation rate and size, and between
linewidth and size, he got the cloud core mass as a function of the
turbulent linewidth. The distribution function for cloud core mass then
followed from a distribution function for linewidth, which came from a
theory for the time-dependent deceleration of wind-driven bubbles in a
cloud.  The star mass is then taken to be proportional to the cloud
core mass.  Silk (1995) made several comments that protostellar
outflows limit the mass of a star, but this assumption was not
present in any of the theory in his paper, nor in the IMF that
resulted, particularly considering that a fixed fraction of the cloud
core mass was assumed to go into the star. This was unlike the results
of the wind-limited accretion models proposed by Nakano, Hasegawa, \&
Norman (1995) and Adams \& Fatuzzo (1996).

One of the most popular and persistent methods for determining the mass
of a star has been with a combination of cloud fragmentation,
accretion, and clump collisions. Larson (1978) showed with
three-dimensional N-body experiments that gas clouds fragment
hierarchically by self-gravity, and the fragments accrete material in
competition with each other. The resulting mass distribution was
modeled after a fractal, which was a remarkably prescient concept in
astronomy considering that Mandelbrot (1977) began to popularize his
fractal geometry only a year earlier.  Larson showed that the fractal
mass function would have a slope of about $x=1$, in reasonable
agreement with observations. Larson also reasoned that the lower mass
limit to a star is the thermal Jeans mass in the cloud because smaller
condensations are not likely to form in the initial collapse. This
experiment was an important departure from standard fragmentation
models of the time, since it was previously believed, following Hoyle
(1953) and Hunter (1962), that fragmentation decreased the Jeans mass,
leading to ever more fragmentation. In Larson's result, the number of
clumps formed was about equal to the number of Jeans masses in the
initial cloud, with no subfragmentation into smaller pieces. This
result was seconded by Tohline (1980), but investigated again by Silk
(1982), who concluded that initial cloud fragments, particularly
elongated fragments, formed by dynamical collapse could in fact
fragment again (see also Bonnell \& Bastien 1993, Burkert et al. 1997).

Silk (1977) considered a different picture of cloud fragmentation,
using a thermal Jeans mass that increased with time as a result of
heating from the stars that already formed. His model of fragmentation
was simpler than Larson's, so there was no built-in, power-law mass
spectrum from the fragmentation process itself. The power law in Silk's
model came from the time increase in the thermal Jeans mass and the
identification of this mass with the stellar mass at all times. Yoshii
\& Saio (1985) followed this model by considering the additional
influence of coalescence among the opacity-limited fragments. They
showed that the opaque fragments are usually so small that coalescence
is unimportant; thus fragmentation and heating alone determine the IMF
as a time sequence of increasing thermal Jeans masses, as in the Silk
(1977) model. Bastien (1981) obtained a different result by using the
Jeans length to determine the fragment collision cross section. The
initial Jeans length is much larger than the size of the opaque
fragments considered by Yoshii \& Saio (1985), so Bastien (1981) found
that fragment collisions were important. Subsequent work by Lejeune \&
Bastien (1986), and Allen \& Bastien (1995, 1996) considered
time-dependent coalescence, and concluded again that both fragmentation
and coalescence were important, with coalescence dominating the
formation of massive stars.

Price \& Podsiadlowski (1995) used protostar collisions in a different
way. They proposed that stars grow by accretion at some more-or-less
uniform rate, and this accretion stops when two protostars collide,
disrupting the gas reservoirs from each. The final stellar mass was
then determined by the product of the accretion rate and the time
interval between collisions. The IMF was built up over time as the
final stellar mass decreased in the presence of an increasing collision
rate that resulted from the continuous formation of more and more
protostars.

Another coagulation model was proposed by Murray \& Lin (1996).  They
suggested that fragmentation driven by thermal instabilities leads to
clumps that fall in the potential well of a cloud, after which they
collide and coalesce with other clumps that have fallen too. A star
forms when the clump mass exceeds the thermal Jeans mass, although
further coalescence can increase that mass afterwards. The power-law
distribution of masses follows from the coagulation process, in a
manner similar to that proposed by Silk \& Takahashi (1979). There is
no assumption about wind-limited accretion here; the star mass is
identified with the core mass directly.  An interesting aspect of their
model is that they considered positionally correlated velocities, as in
a turbulent fluid, but they showed this had no important effect on the
model IMF.

Numerical SPH simulations of an evolving IMF made of accreting,
pseudo-star particle clumps was done by Bonnell et al. (1997). The
advantages of numerical solutions like this is that they can treat well
the competition for gas among all the nucleated centers. They found
that nucleating centers originally close to the center of the whole
cloud grow faster and to larger masses because of the larger gas
density there.

\subsection{An IMF from Random Selection of Mass in Hierarchical Clouds}

A different class of theory considers only the statistical aspects of
the IMF. The Hoyle (1953) picture of fragmentation led naturally to
these ideas, because successive fragmentation produces hierarchical
structure and power law mass spectra regardless of many physical
details.  Larson (1973) considered a modification to these ideas by
proposing that only part of each fragment undergoes further
fragmentation, and got from this a log-normal mass distribution instead
of a power-law.  When Miller \& Scalo (1979) proposed that the IMF
actually had a log-normal form, the statistical implications of this
were developed further by Elmegreen \& Mathieu (1983), Zinnecker
(1984), and Elmegreen (1985), who showed that a variety of processes,
acting together, would combine to make a log-normal function,
regardless of the distribution functions resulting from each separate
process.  This point was made again more recently by Adams \& Fatuzzo
(1996).

The most recent development in this area has been by Elmegreen (1997a,
1998), who considered the random selection of masses from a
hierarchically structured cloud. These papers depart from the usual
scenarios by asserting that hierarchical cloud structure is independent
of star formation -- that it is set up long before star formation
begins in the diffuse cloud stage, and then continuously reestablished
during the molecular cloud stage as a result of supersonic turbulence
compression. There is no gravitationally-driven fragmentation at all,
and no significant clump coalescence. In fact, the clumps in this model
need not even be constant objects, able to move around and coalesce;
they can be ever-changing gas compressions and wavepackets in the
chaotic turbulent flow. This model is motivated by the pervasive
appearance of hierarchical structure and spatially-correlated
velocities in interstellar clouds. These features are reminiscent of
structures and flows in laboratory turbulence (e.g., see reviews in
Sreenivasan 1991, and Falgarone \& Phillips 1991).

The basic point of the Elmegreen (1997a, 1998) model is that
essentially all local star formation processes that have been
considered in detail have a time scale for evolution that varies
approximately as the inverse square root of local density, including
contraction or collapse from self-gravity, turbulence compression,
magnetic diffusion in virialized cores, and coalescence at each level
in the hierarchy. This means that in a model like this, stars will
appear here and there, randomly, alone or with neighbors, at a rate
that depends almost exclusively on the local square root of density. On
a large scale, this implies that dense clouds in high pressure regions
of galaxies will form all of their stars quickly, while low density
clouds in low pressure regions will take much more time. On a small
scale, within any one cloud, it means that different parts of the
hierarchy form stars at different times. The density always increases
at lower levels in a hierarchical structure, and this is where the
clumps, contained by other clumps on larger scales, are small
and have low mass. Thus the low mass pieces in a cloud are likely to
form stars first. As a result, there is slightly less mass for other
stars that form later in the same or higher levels. This competition
for mass tends to steepen the power in the power-law mass function by
several tenths, i.e., from $x=1$ to $x=1.35$, as in the
Salpeter IMF.

Perhaps a more important difference in this new model is the way in
which cloud structure is measured. Typically, the power law index for
molecular and diffuse cloud structure is determined from the mass
distribution of separate clumps, resolved in large-scale surveys by
various telescopes and then separated into discrete objects either by
eye or by computer algorithms that do about the same thing as the eye.
These mass distributions are always much flatter than the IMF, having
slopes $\alpha$ in the range from 1.5 to 1.8, when the IMF has a slope
of about $1+x=2.35$. To understand the IMF, however, we have to try to
view interstellar clouds from the perspective of a forming star. A star
does not care about the clumps that our telescopes resolve and our eyes
choose to label as discrete, but only about the general distribution 
and motion of
gas in all forms. In pre-star-forming clouds, where this structure is
first established, it is largely hierarchical, from scales that are
much smaller than mm-wave telescopes can resolve, up to perhaps a galactic
scale height.

Hierarchical means that most of the small clouds are contained inside
larger clouds, and most small stellar groupings are inside larger
stellar groupings (see review in Elmegreen \& Efremov 1998).  For
example, Efremov (1995) and colleagues have estimated that 90\% of the
OB associations in the Milky Way (Efremov \& Sitnik 1988), M31
(Efremov, Ivanov, \& Nikolov 1987; Battinelli 1991, 1992; Magnier et
al. 1993; Battinelli, Efremov \& Magnier 1996), M33 (Ivanov 1987,
1992), and the LMC (Feitzinger \& Braunsfurth 1984) are inside larger star
complexes. Scalo (1985) has reviewed hierarchical structure for
interstellar clouds. It appears as if the hierarchical embedding of
interstellar and young stellar structures is nearly all-inclusive.

Emission line surveys leading to ``discrete'' clump masses do not
consider this aspect of cloud structure. If two clumps are close
together and part of a larger structure, the algorithms call them two
separate objects, and do not tabulate the larger ``object'' that
contains them. For this reason, all emission line or extinction surveys
catalogue clouds that are within a factor of $\sim3$ to 10 of the
angular resolution, regardless of the cloud distance or the wavelength
of the observation. This factor of $3-10$ for recognized clump size
corresponds to a factor of $\sim100$ for cloud mass (which scales as
the size squared or cubed in CO surveys), so the mass functions look
reasonably well sampled, but in fact only a small part of the cloud
structure is included. Everything smaller than the telescope resolution
is not seen, and everything larger than about 3 to 10 resolution
elements is subdivided into its component parts and not called a
separate object. This is how every cloud or clump mass function has
been evaluated since the beginning of this exercise (i.e., since before
Field \& Saslaw 1965).

The Elmegreen (1997a, 1998) model takes a different point of view. It
considers the structure on all scales and asks for the probability that
any particular mass is chosen from anywhere in the whole hierarchy.
This is presumably what happens as a result of the combination of
physical processes that leads to star formation in a real cloud:
because of the self-similar nature of turbulent flows, each level in
the hierarchy looks the same as any other level (for masses above the
thermal Jeans mass), so each choice of level for a star-formation event
would have the same likelihood as any other choice, modulated only by
the density-dependent local evolution rate discussed in the previous
paragraph. Aside from this variable rate, the instantaneous
distribution of masses for all structures, and the instantaneous
probability of selecting any particular mass, is proportional to
$M^{-2}$ for linear intervals in mass (i.e., $\alpha=2$ in the notation
of the clump spectra given above). Such a distribution for hierarchical
structure was also recognized by Larson (1978) and Fleck (1996). The
result gives a steeper instantaneous mass function than emission line
or extinction surveys, but there is no conservation of mass as in these
standard surveys (i.e., the sum of the masses of all possible
structures is larger than the mass of the whole cloud, because nearly
all of the structures are contained in other structures, and so are
multiply counted).  This way of viewing cloud structure seems to be
closer to what a real star-forming process ``sees'' before it actually
begins the sequence of events that makes a star. The density dependent
rate of star formation then steepens the mass function from $M^{-2}$
to $M^{-2.35}$. This result is then identified
with the Salpeter IMF.

For the purposes of understanding the IMF, there are probably better
ways to measure cloud structures than with clump-finding algorithms.
The structure on the edges of clouds has been analyzed in terms of
fractals, rather than clumps, for many years (Beech 1987; Bazell \&
D\'esert 1988; Scalo 1990; Dickman, Horvath, \& Margulis 1990;
Falgarone, Phillips, \& Walker 1991; Zimmermann \& Stutzki 1992, 1993;
Henriksen, 1991; Hetem \& Lepine 1993; Vogelaar \& Wakker 1994;
Pfenniger \& Combes 1994).

For the structures inside clouds, Stutzki et al. (1998) showed that
Fourier transform power spectra of emission line intensity scans across
molecular clouds gave power laws, and concluded that the internal
structure was scale-invariant, as in a fractal.  They reproduced this
structure with a random fractal-generation model.  Other models that
made cloud fractals were in Hetem \& Lepine (1993) and Elmegreen
(1997b). The concept that the cloud mass distribution function is the
result of fractal structure, presumably generated by turbulence, began
with papers by Fleck (1996), Elmegreen \& Falgarone (1996), and Stutzki
et al. (1998).  This seems to be a more reasonable explanation than the
older collisional-build-up models of clump structure, particularly
since off-center supersonic collisions between clumps should not be
sticky (Scalo \& Pumphrey 1982; Kimura \& Tosa 1996; Fujimoto, \& Kumai
1997).

The probability of selecting structures for star formation in
hierarchical clouds seems to show up more directly in the distribution
of masses for open clusters. When a bound cluster forms, a high
fraction of the gas mass has to go into stars (see Verschueren 1990 and
references therein), so the cluster mass distribution should reflect
the mass distribution of cloud structures pretty well. In this case,
the result is obviously independent of telescope resolution or 
cloud-clump 
recognition bias, because clusters are seen optically, even at
great distances. Also, the range of masses for clusters can be
reasonably large, exceeding a factor of 100, so a power law can be
measured fairly well if it exists. Cluster selection suffers from
various other biases, however, such as extinction, age limitations, and
a loss of low-mass clusters with increasing distance. Nevertheless,
there are some studies that get around these biases, and they confirm
the expected result. In two samples of nearby clusters, each with
calibrated masses, Battinelli et al.  (1994) found power law mass
functions with slopes of $-2.13\pm0.15$ and $-2.04\pm0.11$. These
clusters are close enough to the Sun to be relatively free of distance
and extinction effects. Also, in the LMC, where $\sim600$ clusters are
catalogued (Bica et al. 1996), calibrated photometrically for age
(Girardi et al 1995), and all at about the same distance, Elmegreen \&
Efremov (1997) found $M^{-2}dM$ mass functions for discrete age groups
(e.g., $10^8-10^9$ yrs; dividing the clusters into age groups is
important when there are no direct measures of cluster mass, because
the conversion from luminosity to mass depends on age).

The IMF should be steeper than the cluster mass function because of the
density-dependent formation rate of individual stars, and the
competition for mass in the IMF. These effects are not important for
clusters. Open clusters take a high fraction of all the gas mass
originally available to them (or else they would not have ended up
bound), and the rate at which they do this is not important for their
final masses. In the case of stars, new objects forming at one level
steal mass away from the objects that form later at a higher level. But
clusters do not do this:  those which form inside other clusters in the
general hierarchy can just merge together to make a single cluster of
higher mass, appropriate to the higher level in the hierarchy. And if
they do not merge, then they stay with their original masses, which are
appropriate for the levels in which they form.

Other IMF models based on fractal or hierarchical cloud structure were
developed earlier by Henriksen (1986, 1991) and Larson (1992).
Henriksen used the size, $L$, distribution function of structures in a
fractal of dimension $D$, given as $n(L)d\log L\propto L^{-D}d\log L$
by Mandelbrot (1983), and assumed that the density $\rho$ varied with
size too, not as in a fractal (which would be $\rho\propto L^{D-3}$)
but as actually observed in self-gravitating clouds, namely
$\rho\propto L^{-1}$. This gave a mass function for cloud structure
that agreed well with observations if $D\sim2.7$. Note that if
Henriksen used the density dependence for a fractal, in a
self-consistent manner with the size distribution, then the resulting
mass function would have been independent of the fractal dimension,
$M^{-2}dM$, as discussed above for hierarchical clouds in general.
Larson (1992) started again with the size distribution for a fractal,
and assumed that final stellar mass is directly proportional to the
linear size of the cloud structure, suggesting that such structures
were filamentary anyway. Then the slope of the IMF in linear intervals
became $1+x=1+D\sim3.3$ for $D=2.3$, as implied by the fractal
dimension of cloud perimeters, namely $D_p\sim1.3$, added to 1 to
account for the higher dimension of the surface. This IMF, with a slope
of $1+x=3.3$, is significantly steeper than the Salpeter IMF, with a
slope of $1+x=2.3$, but Larson suggested that maybe stars form in
subparts of clouds where $D$ is smaller than 2.3, or perhaps larger
structures have larger temperatures, which would break the assumed
linear relationship between star mass and scale size.

These models differ from the Elmegreen (1997a, 1998) model, even though
both employ ``fractal'' cloud structure, because the latter uses general
cloud shapes, not necessarily filaments or sheets, random sampling from
all levels in the hierarchy of structures (giving $M^{-2}dM$ from
sampling alone), a density-dependent rate that steepens the IMF by
preferred sampling at lower mass (i.e., steepening the result to
$M^{-2-0.5+1.5/D}\sim M^{-2.15}$), and a competition for mass, which
steepens the IMF further, to $M^{-2.35}dM$, as in the Salpeter IMF. With
these different assumptions, the IMF in the Elmegreen model hardly
depends on the fractal dimension at all; it enters into the exponent of
the IMF power law as approximately $0.5-1.5/D$, which is a very small
increment for $D$ in the likely range of 2 to 3.

\subsection{Theories for The Lower Stellar Mass Limit}
\label{sect:theorylm}

Understanding the lower mass limit to star formation is an important
part of any IMF theory. Various assumptions about this have already
been mentioned in the review of theories presented above. Three views
on this subject seem to pervade most of them. In the wind-limited
accretion models, the assumption is made that accretion onto a dense
core continues until a protostellar wind develops, which presumably
follows the onset of deuterium burning (Shu et al. 1987). Thus low mass
objects do not stay that way very long, they just keep accreting
until they become active protostars. In most of the older
fragmentation-coalescence models, the minimum mass is the mass of a
gravitationally bound fragment that is optically thick. In the various
scale-free models, based on gravitationally-driven fragmentation or
turbulence, the minimum mass is usually the thermal Jeans mass,
particularly in the Larson (1992) and Elmegreen (1997a, 1998) models.
This assumption is made because, without further fragmentation during
the collapse itself, the thermal Jeans mass is the smallest structure
that can ever become self-gravitating, even in the presence of
turbulence and magnetic fields.

The first and the third of these mass limits equals about the observed
limit, $\sim0.1$ M$_\odot$. In the first case, it is self-defining:
objects accrete until they become star-like, and then they stop
accreting, so naturally the lowest possible mass is the mass of a star.
This concept has the powerful advantage over the others that the basic
star mass, and perhaps even the IMF itself, can be the same everywhere,
independent of cloud conditions. In the third case, the minimum
theoretical mass just happens to equal the observed minimum in standard
cloud conditions. An expression for this mass is the Bonner-Ebert
critical mass, \begin{equation}
M_{BE}=0.35\left({{T}\over{10K}}\right)^{2}\left({{P}\over
{10^6\;k_B}}\right)^{-1/2}\;\;{\rm M}_\odot.       
\label{eq:mj}           
\end{equation}
Here we have normalized the result to standard cloud conditions, taking
the pressure inside a cloud equal to the typical self-gravitating core
pressure, not the boundary pressure (which is sometimes lower by a
factor of $\sim30$). Our use of this form for the thermal Jeans mass is
motivated by the near constancy of $T$ and total $P$ in molecular
clouds, according to the scaling relations (Larson 1981).  This is much
more sensible than writing the thermal Jeans mass in terms of $T$ and
density, for example, when the density varies by several orders of
magnitude inside a cloud. Our use of the total pressure for $P$ also
allows for a decrease of Jeans mass in compressive flows, as 
in Hunter \& Fleck (1982).

Nevertheless, another
theory of the IMF
in which all of the stellar masses have the local thermal
Jeans mass has been advanced by Padoan, Nordlund, \&
Jones (1997), 
with variations in mass entirely the result of variations
in the density in a turbulent fluid.  In this model, the high mass
stars form in the low density regions, because the Jeans mass is high
there for a given temperature.

The second minimum fragment mass discussed above, from the opacity
limit of a self-gravitating condensation, is typically much smaller
than the minimum star mass, so coalescence has to bring the star mass
up to the observed value in these models.

Larson has looked for signatures of the thermal Jeans mass in several
ways, predicting that it should vary with cloud temperature (Larson
1985, 1992). In Larson (1992), he suggested that such variations were
observed by Myers (1991) in different molecular clouds. Larson
(1995) also suggested that the length scale at a break in the
separation distribution for T Tauri stars was the Jeans length (see
also Simon 1997; but see Bate, Clarke, \& McCaughrean 1998 
and Nakajima et al. 1998 for other interpretations.)

Elmegreen (1997a, 1998) considered $M_{BE}$ from a different point of
view, stating that most star-forming regions in normal galaxies
actually do have nearly constant values of $T^2/P^{1/2}$, with small
variations in this quantity possibly leading to the observed variations
in cluster IMFs at low mass (see Section \ref{sect:intro} above). This
constancy is a result of the numerator in $M_{BE}$ being approximately
proportional to the cooling rate per unit mass of molecular gas
(Neufeld, Lepp, \& Melnick 1995), and the denominator being
approximately proportional to the local surface density of stars and
gas in the galaxy, which determine the heating rate per unit mass from
cosmic rays and background starlight. As long as cooling roughly
balances background heating, this ratio will be about constant. Other,
more local variations in $T$ and $P$ tend to cancel out when combined
as the quantity $T^2/P^{1/2}$.  For example, when $P$ goes up in a
triggered region next to a previous generation of star-formation, $T$
generally goes up too because of the enhanced
radiation field. Conversely, when $P$ is low, as in low-mass clouds
like Taurus, then $T$ is low as well.

Elmegreen (1997a, 1998) also points out that the lower mass limit for
stars actually {\it does} change, perhaps by a factor of 10 in extreme
starburst regions (see Sect. \ref{sect:intro} above) where the
average $T$ can be
much larger than in local molecular clouds, e.g., 100 K instead of 10 K
(cf. Aalto et al. 1995), and the pressure much larger too (e.g.,
$\times10^4$). But this variation should be rare in normal
galaxies because of feedback
processes in the interstellar medium that tend to connect young stellar
activity, interstellar pressure, and gas temperature. If we consider
$T$ to be a measure of the rate of cooling per unit cloud mass, then
$T$ in a star-forming region will be high when there is an unusually
large embedded or nearby luminosity of stars per unit gas mass. This
occurs when the efficiency for star formation is high in a high mass
cluster (the cluster has to be fairly high mass to contain the luminous
stars that dominate the total radiation field). A high efficiency in a
low mass cluster would not necessarily increase $M_{BE}$ because low
mass clusters are not very luminous per unit mass. A low efficiency in
a high mass cluster, leading to an OB association, might not change
$M_{BE}$ much either, because the radiation luminosity per unit gas
mass is not particularly large there either. Thus the formation of
globular clusters in regions with only moderately large pressures, not
the enormous pressures typical of early galactic halos, might be
elevate $M_{BE}$, but this conclusion is very uncertain.

\begin{figure}
\vspace {2.8in}
\includegraphics{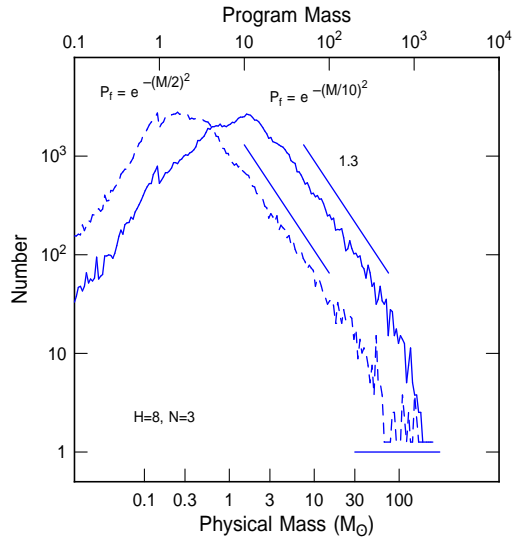}
\caption{Two model IMFs with different lower mass limits, $M_{BE}$. The
entire IMF shifts to the right when $M_{BE}$ increases because the
power law portion always has the same slope. The fall-off at high
mass is the result of computational limitations.
The models have $H=8$ levels in the hierarchy with
$N=3$ subclumps per clump, on average.} 
\label{fig:mj}
\end{figure}   

Figure \ref{fig:mj} shows two numerical experiments that made model
IMFs from random selections in a hierarchically structured gas cloud,
using a rate of selection proportional to the square root of the gas
density, as discussed above (Elmegreen 1997a). The models differ only
in the value of $M_{BE}$, which enters into the probability for failure
to form a star, taken to be $P_f=e^{-\left(M/M_{BE}\right)^2}$ in these
cases ($M_{BE}$ in the figure is written in terms of the
computer-program mass, which is scaled to a physical mass after
multiplication by $0.15$.  The IMFs are
identical except for the larger value of the lower mass limit in the
case with high $M_{BE}$.

\subsection{A Combination of Theories?}

The two main contenders for the lower mass limit of star formation,
wind-regulation and the thermal Jeans mass, have obvious differences in
predicted behaviors of the shape of the IMF at low mass. If this shape
varies a lot, with some clusters having clearly higher lower mass
limits than other clusters, then wind self-regulation might be ruled
out.  Moreover, if cluster IMFs in regions with large turnover masses
also have larger pre-stellar $M_{BE}$, then the thermal Jeans mass
would seem to be important.

Curiously, both of these models, the wind-regulated IMF, and the
$M_{BE}$--fractal cloud model, have their strong points where the other
is weakest.  The wind-regulated IMF model seems to have
difficulty explaining the nearly universal slope of the IMF without
some sensitivity to cloud parameters, yet it seems certain that wind
self-regulation must play some role in determining the mass that gets
onto a star, at least near the lower mass limit for star formation.
Similarly, models that rely on $M_{BE}$ have an uncomfortable
susceptibility to changes in cloud properties near the lower mass
limit, but get the Salpeter slope above this limit with essentially no
free parameters.  Perhaps a combination of the two models would be
better.

There are several ways this could occur. First, the fractal model
supposes that a star's mass can be identified with the mass of some
structure in a turbulent cloud, but this is obviously too simple --
there is a lot that can happen during star formation that will vary the
fraction of the cloud mass that gets into a star.  If this fraction is
randomly distributed, then the same IMF slope and scatter about that
slope result, as shown in connection with figure \ref{fig:scatter}
below, which is discussed in more detail in Elmegreen (1998). If it is
not randomly distributed, then the power-law slope could change.  But
the entire power-law slope is not likely to come from wind-regulation,
that would give the slope too great a sensitivity to cloud properties
and the randomness of wind-clearing.  This means that, perhaps to
within a factor of 2 to 5, star mass should really be identified with
the mass of the clump in which it forms, i.e., that bigger stars really
do form in bigger clumps (Casoli et al. 1986; Myers, Ladd, \& Fuller
1991; Myers \& Fuller 1993; Pound \& Blitz 1995; Motte, 
Andr\'e, \& Neri 1998; not to be confused
with the statistical sampling statement discussed in section
\ref{sect:intro}, that bigger stars form in bigger {\it whole}
clouds).  This factor of 2 to 5 may contain all the detailed physics of
wind clearing.

Protostellar winds should also play some role in regulating the
smallest stars that form.   According to the models by Shu and
collaborators, as reviewed in Shu et al. (1987), the onset of the wind
is triggered by deuterium burning in a young, pre-stellar object.  If
the object is too small, there will be no wind, and presumably
accretion will continue until deuterium burning begins.

This concept combines well with the arguments based on a thermal Jean
mass.  For example, if $M_{BE}$ is much smaller than the deuterium
burning limit, then a large number of small brown dwarfs or Jupiters
should form, and these could either leave the cloud in that state,
forever drifting as brown dwarfs, or they could accrete more mass over
time and end up finally able to start a wind. In the latter case, the
effective lower mass limit to the IMF would be the deuterium burning
mass, not $M_{BE}$, and the larger masses would follow the fractal mass
distribution as before, giving the Salpeter function.  On the other
hand, if there are a lot of free brown dwarfs, which does not appear to
be the case in the Solar vicinity (cf. Sect. \ref{sect:intro}), and if
there are also a lot of regions where $M_{BE}$ is low, which
does not appear to be the case locally either, then a model with
accretion up to a wind-limited mass would not seem to work.

If $M_{BE}$ is much larger than the deuterium burning limit, then
large, self-gravitating clumps will form in the cloud, but the rapid
onset of a wind in these clouds might limit the gas mass that actually
gets into the star to only a small fraction of what is available. Then
all of the excess mass between the deuterium limit and M$_{BE}$ would get
dispersed back into the cloud for recycling into other stars, and the
lower mass limit would be the deuterium limit.  In this second case,
the mass fraction that goes into each star will be small, but if it is
about the same fraction for all clumps, then the higher mass stars will
still map out the fractal properties of the cloud and produce the
Salpeter IMF.  Again the details of how the IMF gets established in a
cloud depend strongly on unknown processes during the final collapse
and dispersal phase of star formation, but the basic shape of the IMF
could be relatively independent of these details.

\subsection{Reflections on the Various Theories of the IMF}

There have been some fundamental changes lately in how we view molecular
clouds and star formation, and these changes inevitably affect the
various models for the IMF. 

For example, we are beginning to think that star formation is
relatively rapid in cloud cores, occurring in only one or two
crossing times regardless of scale (Elmegreen \& Efremov 1996; Efremov
\& Elmegreen 1998). This means that we cannot generally wait for
magnetic diffusion to occur as a cloud core slowly accretes across
field lines. Diffusion typically takes about 10 free fall times for
cosmic ray ionization (Shu et al. 1987), longer for ionization by
starlight (Myers \& Khersonsky 1995), and longer still if the gas is
clumpy (Elmegreen \& Combes 1992). Indeed, Nakano (1998) suggested that
all star-forming cloud cores are magnetically {\it super}critical;
i.e., they would collapse dynamically across the field lines if
turbulent motions were not present.

There are two important implications of this change in thinking, if
correct. First, there would no longer be a strong motivation for IMF
theories that consider bimodel star formation in the sense that low mass
stars come from subcritical cores, and high mass stars come from
magnetically supercritical cores (Lizano \& Shu 1988). The IMF is so
remarkably uniform anyway, it does not seem possible to have widely
different modes of star formation at high and low mass. Any change from
low mass dominance to high mass dominance in the IMF could more easily
be accommodated by an upward shift in the thermal Jeans mass (cf.
Section \ref{sect:theorylm}).

Secondly, this time scale for star formation may be too short to allow
multiple interactions between protostellar clumps. This would then
rule out a broad class of coalescence and slow accretion models. Such
coalescence is unlikely anyway for clumps that move supersonically
relative to each other; they will fragment or disperse upon collision,
rather than stick together. Clump collisions may trigger star formation
when they occur (Bhattal et al. 1998), but multiple collisions are
probably not the source of the cloud or star mass distributions. 

Another observation that suggests the same thing is the extremely high
stellar densities in young globular clusters, which form today in
starburst regions. As suggested in Section \ref{sect:intro}, these
globulars look like they will evolve into objects similar to our Milky
Way's halo globulars, and the old Milky Way globulars have an
apparently normal IMF. If coalescence is important in local ``normal''
embedded clusters, it would seem to be devastating in globular
clusters, where the stellar density can be $10^4$ times higher.
Conversely, if these globulars have normal IMFs, then protostellar
coalescence must be unimportant nearly everywhere.   A similar
constraint would come from the IMF in small regions, where only a few
stars form. There, protostar interactions are not likely to be important
either.

A second major shift in our view of molecular clouds is that most
astronomers now think that pre-stellar structure comes from turbulence,
not gravitationally-driven fragmentation. This is because the same
character of structure is observed in both self-gravitating and
non-self-gravitating clouds, and both types of clouds have correlated
motions reminiscent of laboratory turbulence. A related point is that the
cloud cores in which clusters form are not obviously collapsing as a
whole: there are no inverse P-Cygni profiles indicative of collapse
motions for whole cores, as there are in some individual protostars.
The stars are apparently forming inside more-or-less stable cloud
cores. Once again, the general collapse models seem untenable. 

The presence of supersonic turbulence also seems to rule out models
involving thermal instabilities. Turbulent motions in cluster-forming
cores are generally much larger than thermal, and so the dominant forces
that structure the gas are the turbulent forces, not the thermal. This
means that turbulence causes the observed structures in pre-stellar
clouds, not thermal instabilities. Thermal processes are probably
important on the smallest scales, but these are at and below $M_{BE}$.

Turbulence in pre-stellar clouds also appears to generate structure
that is much lower in mass than even the smallest stars, perhaps as low
as $10^{-4}$ M$_\odot$ (Heithausen et al. 1998; Kramer et al. 1998). 
This means that star
formation occurs in the middle range of all cloud structures, not at
the bottom end. It also means that, while turbulence may generate a
cascade of structures down to
smaller and smaller scales, the end result of this cascade is not the
formation of a star. Turbulence makes cloud structures independent of
star formation. Self-gravity, sonic wave generation, magnetic
reconnection, and many other aspects of interstellar cloud dynamics are
likely to be important too by the time star formation begins.

\subsection{Implications of a Nearly Uniform IMF}

The remarkable near-uniformity of the IMF in regions with a wide range of
stellar densities, cluster masses, metallicities, ages, and galactic
types strongly suggests there is a {\it unifying process in star
formation} that makes the relative probability of forming various
masses somewhat independent of physical parameters. For example, the
IMF seems independent of how star formation begins, i.e.,
whether the region is triggered by some external pressure or forms
quiescently (e.g. see Parker et al. 1992 and the contrary opinion by
Oey \& Massey 1995), independent of the general cloud shape (shell,
layer, filament, etc.), magnetic field strength, ionization level,
presence of other stars, and so on. This is remarkable because these
physical variables do control the {\it rate} of star formation in some
regions, and they certainly control {\it where and when} star formation
begins. They just do not appear to influence the IMF.

The near-uniformity of the {\it slope} of the power law
also implies several other things. (1) There is a {\it
universal aspect of cloud structure and evolution that produces the
power-law portion of the IMF independently of the temperature,
pressure, and metallicity.} This means that the power-law probably does
not depend on the protostellar collapse process, e.g., on the
fragmentation or collapse of isothermal clumps, the accretion of gas in
thermal equilibrium, or the thermal Jeans mass. For all of these
processes, thermal temperature and magnetic diffusion are probably
important at some stage (see reviews in Nakano 1984; Mouschovias 1991;
Shu, Adams, \& Lizano 1987).  (2) The IMF power-law is also likely to
be a {\it highly reduced average over many physical processes that
either all give about the same result, or are combined so finely in
every region that the same proportion of each process is always
present.}

The variations in slope of the IMF from region to region (Scalo 1998)
seem at first to suggest something different, that the IMF is not in
fact uniform but depends sensitively on physical conditions.  But are
these variations statistically significant?  Rarely do studies of
cluster IMFs contain more than several hundred stars.

\begin{figure}
\vspace {3.1in}
\includegraphics{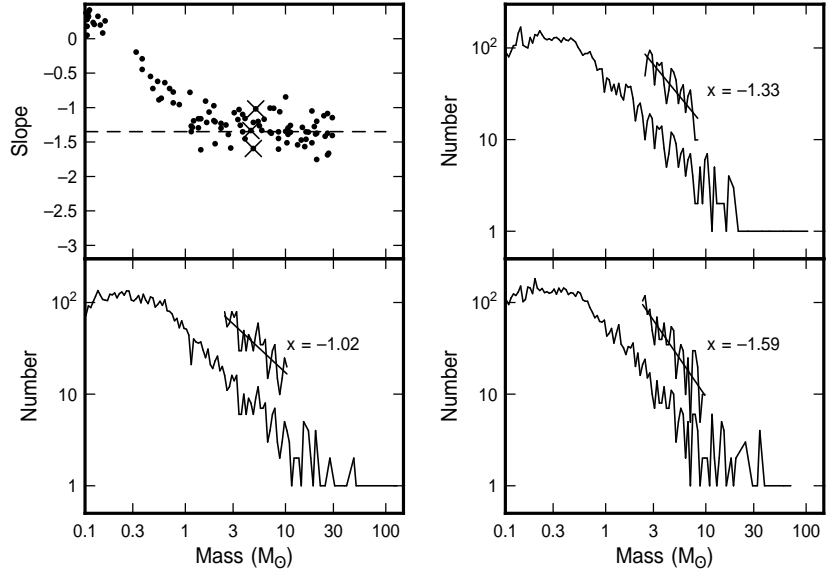}
\caption{(top left) IMF slopes in 100 models, plotted as a function of the average
logarithm of the mass. Each IMF slope is fit using
200 stars, with the largest star in the fit taken to be smaller
by 20 filled mass bins than the largest star produced in the model.
The Salpeter value of $-1.35$ is shown by a dashed line.  The three
values indicated by crosses have their complete IMFs shown in
the other panels. The fitted portions of these IMFs are indicated
by the offsets.} 
\label{fig:fluctuations}
\end{figure}   

Figure \ref{fig:fluctuations} shows sample IMFs from the model of
random selection in a hierarchical cloud (Elmegreen 1998). The panel in
the upper left shows IMF slopes for 100 different models, each with a
different number of stars and therefore different upper mass limit. In
all cases, the range of stars chosen for the power-law slope
determination is such that there are 200 stars in the fit, starting
from the histogram bin that is twenty filled bins (a factor of 3 in
mass) away from the most massive star, and going to lower mass bins.
This starting point for the fit ensures that there are enough stars to
get an accurate slope, and it coincides with an astronomer's decision
to avoid fitting the IMF to the scarce few highest-mass stars in the
cluster.  We pick a different number of total stars for each IMF, and
use only the highest mass parts of each model, because this is what an
astronomer would do as well.  Generally an observer catalogues only the
brightest stars in a cluster and misses the low mass members. If a
cluster has an IMF made only from low mass stars, then that cluster is
generally so small that there are no high mass stars at all.

Evidently the IMF slope varies a lot around the Salpeter value, shown
in the figure by the dashed line at $x=1.35$. This average slope is
nearly independent of physical cloud properties in this model (see the
discussion in section \ref{sect:theory}).  The other
three panels in figure \ref{fig:fluctuations} show particular IMFs and
the mass ranges and fits used for the ``X'' marks in the top left
panel.

\begin{figure}
\vspace {2.15in}
\includegraphics{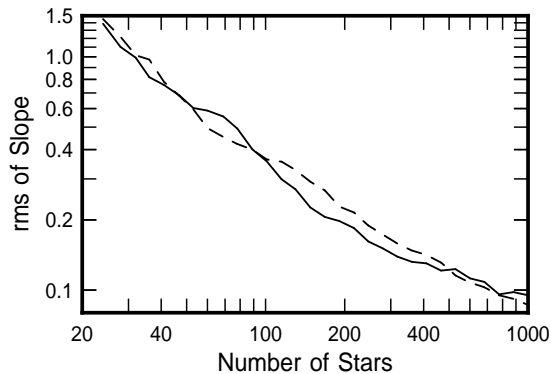}
\caption{
The rms deviation in the fitted slopes of the model IMFs for
the 1-10 M$_\odot$ range, plotted versus the number of stars
that are included in the fit.  The solid line is for
the case shown in Fig. 4, and the dashed line is for another
100 independent IMFs that have an additional randomness in the
ratio of the star mass to the clump mass. 
} 
\label{fig:scatter}
\end{figure}   

Figure \ref{fig:scatter} shows the rms deviations in the IMF slopes
around the mean values for power-law fits that include different
numbers of stars in the same 100 IMFs (from Elmegreen 1998).  Each
plotted value, connected by a smooth line, represents the rms
deviation for the 100 models, calculated in the mass range from 1 to 10
M$_\odot$ (the characteristic physical mass in the model comes from
$M_{BE}$, which is taken to be 0.35M$_\odot$ from equation
\ref{eq:mj}).  The rms deviation around the Salpeter slope decreases
systematically as more and more stars are included in the fit.  The
deviation can be as high as $\pm0.5$ when only 80 stars are used,
decreasing to $\sim0.1$ when 1000 stars are used. This implies that an
IMF calculated with, say, 150 stars, and having a slope of $-1.15$
instead of the Salpeter slope of $-1.35$, is in fact statistically
consistent with this Salpeter slope in a ``universal'' IMF.  The
fluctuations found by Scalo (1998) in his review correspond to $\pm0.5$
around the value $x\sim1.35$. These fluctuations could be statistical,
considering the number of stars included in typical IMF surveys and the
likely presence of measurement and star selection errors in the real
data.

The dashed line in figure \ref{fig:scatter} is for another 100 models,
independent of the first, and evaluated with a large ($\times 4$) 
and random
fluctuation in the ratio of the star mass to the clump mass. Such
variations do not affect the average IMF slope or the statistical
fluctuations around it.

\section{Summary}

The IMF has been observed directly and indirectly for many years, with
many different results, but there seems to be some convergence now in
the basic form of the IMF, and also some tantalizing indications that
this form changes a little when the physical conditions for star
formation change a lot. The main observational results were summarized
in section \ref{sect:sum1}.

The theory of the IMF seems to be all over the map, even in recent
years. This variation reflects our ignorance in the processes that
determine a star's mass and in the processes that cause the complex
spatial structures seen in interstellar clouds. The common aspects in
many of these models, including clump and protostellar interactions,
wind erosion of protostellar gas concentrations, minimum masses from the
lack of self-gravity, and hierarchical cloud structure, are all likely
to play some role in generating the IMF. The worry is that the IMF is
such a highly reduced average over many physical processes that each
process does not a have clear signature in the final result. 

In view of the observations, we should perhaps consider ``acceptable''
those models that do not have much sensitivity to either the local
aspects of cloud structure, or the large scale aspects of time and place
in the Universe. This drives the recent appeal towards models based on
common turbulence in one way or another, considering that turbulent
structures are likely to be robust. Unfortunately, we do not understand
compressible MHD turbulence much either. 

There are several key observations that would help. One is the
determination of star formation time scales, always in units of the
cloud or clump crossing time. There are three important timescales: the
rise time of the star formation rate in a cloud, the duration of star
formation, and the decay time scale. If the rise time is less than a
crossing time, and the duration only one or two crossing times, then
IMF models based on multiple clump interactions would seem to be ruled
out.

There is also a need to know the mass dependence of the mass fraction
of clump gas that goes into a star. This might be determined from the
ratio of luminosity to gas mass for Type 0 sources, plotted as a
function of gas mass for the dense cores where the stars are actually
forming.

Theory should tell us what gas structures and evolutionary timescales
are expected from self-gravitating MHD turbulence in the
supersonic-subAlfv\'enic regime of molecular clouds. If turbulence only
makes moving waves and wavepackets, and star formation actually occurs
in such regions, then star formation has to be relatively quick in
terms of the local crossing time. It may be that in the absence of strong
self-gravity, turbulent clumps are transient and amorphous, but when
self-gravity becomes important, the clumps get some integrity and
persist for a relatively long time.

\begin{acknowledgments}
Thanks to Gerald Gilmore for pre-publication copies of reviews from 
the IMF conference that he organized with I. Parry, and S. Ryan. 
\end{acknowledgments}

\end{document}